\documentclass[journal]{IEEEtran}

\ifCLASSINFOpdf
\else

\fi

\usepackage{amsmath}
\usepackage{esint,color}
\usepackage{subfigure}
\usepackage{gensymb}
\usepackage{cite}
\usepackage{graphicx}

\usepackage{tabularx,booktabs}
\newcolumntype{C}{>{\centering\arraybackslash}X} 
\usepackage{lipsum} % filler text

\begin{document}

\bstctlcite{IEEEexample:BSTcontrol}

%
%\title{Directional Modulation by Dynamic Current Distribution of a Single Antenna}
%\title{A Single Port Single Antenna for Wireless Security Via Directional Modulation}
\title{A Dipole Antenna with a Dynamic Balun\\for Wireless Security via Amplitude \\ Based Directional Modulation}
\author{Amer Abu Arisheh,~\IEEEmembership{Graduate Student Member,~IEEE} and Jeffrey A. Nanzer,~\IEEEmembership{Senior Member,~IEEE}
%\thanks{Manuscript received 2021.}
\thanks{This material is based in part upon work supported by the National Science Foundation under grant number 2028736. \textit{(Corresponding author: Jeffrey A. Nanzer)}}
\thanks{The authors are with the Department of Electrical and Computer Engineering, Michigan State University, East Lansing, MI 48824 USA (email: \{abuarish, nanzer\}@msu.edu).}
} % 

\markboth{IEEE}%
{Shell \MakeLowercase{\textit{et al.}}: Bare Demo of IEEEtran.cls for IEEE Journals}

% \thanks{This paper is an expanded version of a conference paper presented at IEEE Int. Symposium on Antennas and Propagation, Atlanta, GA.}




\maketitle


\begin{abstract}

We present a new approach to secure wireless operations using a simple dipole antenna with a dynamic unbalanced feeding structure. By rapidly switching between two states, a dynamic radiation pattern is generated, resulting in directional modulation. The current distribution on the arms of the dipole antenna are made asymmetric by the balun, which changes the relative phase between the feed currents. By rapidly switching between two mirrored states, the relative phase shift changes sign, causing the current distribution to manifest asymmetrically on the arms of the dipole antenna. The resultant far field radiation pattern is therefore asymmetric in both states, but mirrored between the two states. Rapid switching between the two states results in a far-field pattern that is dynamic in amplitude at all angles except for a narrow region of space, which is referred to as the information beam. The dynamic radiation pattern causes additional modulation on any transmitted or received signals, thereby obscuring the information at all angles outside the information beam. The proposed directional modulation technique is separate from both the antenna and the rest of the wireless system, and can thus be implemented in a black box form in wireless communications or sensing systems. We demonstrate the concept in a 1.86 GHz printed dipole antenna, demonstrating the transmission of 256-QAM signals.

%We introduce a single port single antenna for physical layer security via Directional Modulation (DM) technique. We design a printed dipole antenna fed by an imbalanced feeding structure that is switched constantly and rapidly to generate two mirrored states of the antenna. Thus, at the operating frequency of 1.86 GHz, the antenna exhibits fixed field at specific direction(s) and simultaneously dynamic fields elsewhere. Consequently, large modulation orders of QAM are received correctly at only the fixed field direction(s) while are distorted elsewhere. Theoretical analysis is provided to analyze the antenna operation which mainly relies on different relative phases between the two dipole arms, as if it is an array technique achieved in element level. This relative phases generates dynamics by steering the amplitude pattern without significant phase center displacement, and fabricated antenna is tested experimentally in a live DM demonstrator.  The design of the antenna also supports control of information beam direction and its width. To the best of our knowledge, Directional Modulation has never been achieved by a single port single antenna; it is commonly done by arrays or multiport antennas. Our proposed antenna can be easily used in a ``black box" fashion to achieve physical layer security at the element level.

\end{abstract}


% \fi
\IEEEpeerreviewmaketitle

\section{Introduction}
\label{Introduction}

The increasing prevalence of wireless communications in compact, low-power devices for applications like IoT has driven the need for wireless security approaches that are power efficient and able to be implemented at the physical layer. Protecting user information and devices from eavesdroppers, interference, or malicious transmissions requires security at multiple levels, including traditional approaches like encryption. However, physical layer approaches implemented at the antenna or array level have the added potential of direction-dependent security, where the desired information can be transferred to a specific destination while ensuring that transmission and/or reception of information at other directions is challenging or impossible. Directional modulation is a technique where the transferred signal is modulated differently as a function of angle, and can provide direction-dependent security. 

Implementation of directional modulation by antennas and arrays has been studied extensively in recent years, initially focusing on antenna arrays and more recently on multiport antenna structures~\cite{2021,cabrera2022multibeam,zandamela2022directional}. The objective in all cases is to send undistorted signals to an intended receiver (or, equivalently, to receive undistorted information from an intended transmitter), while ensuring that any data transferred at other directions is sufficiently modulated by additional time-varying amplitude and/or phase weights such that the information is unrecoverable. The direction-dependent time-varying modulation manifests in the radiation pattern of the aperture and thus generally has to be implemented at the physical layer by altering the current distribution on the aperture. 
Directional modulation can be obtained in array structures in various ways, such as modulating the amplitudes of some elements~\cite{6544472} or by appropriate phase weighting~\cite{5422702}. Directional modulation in more compact apertures than arrays generally focus on structures that are effectively multiply resonant antennas with several weighted feeding points activated simultaneously~\cite{ding2013vector,2017,narbudowicz2022energy}.

Because the time-varying radiation pattern results from changes in the current distribution on the aperture, implementations of directional modulation on single-antenna structures are challenging, and have only recently been explored in our prior work~\cite{arisheh2023design}.
Specifically, we utilized two mirrored current states of an off-center-fed dipole antenna~\cite{suppressing}, which were excited sequentially at two points on the antenna, yielding an asymmetric current distribution and thus an asymmetric dynamic radiation pattern. While two feed points were used, only one signal feed was necessary.
Off-center-fed dipole antennas exhibit current having both odd and even symmetrical modes, which manifest along with simultaneous common-mode and differential-mode current excitations~\cite{ishii2014analysis}. These antennas have been used for various applications~\cite{prinsloo2013design,prinsloo2014mixed,qin2013beam,xiao2021dipole,li2015novel}, but have not been explored for directional modulation purposes. 
In~\cite{prinsloo2013design,prinsloo2014mixed}, a dipole and monopole were used within same structure to combine both differential and common modes to enhance the coverage area. 
In~\cite{qin2013beam}, a Quasi-Yagi dipole antenna was implemented with a balun feed whose length was controlled by PIN diodes, supporting pattern reconfigurability. 
%Although they discussed the relation of current phase unbalance on dipole and steering for a certain direction, measurements did not purely prove their theory or their balun design method since they used large number of parasitic elements directed toward the desired steered direction and thus they did not isolate the problem; also, they use the antenna in a static manner making it a reconfigurable antenna, not a constantly and rapidly switched antenna. 
In~\cite{xiao2021dipole}, both even and odd modes were excited simultaneously on a dipole fed by a single port using a capacitive coupling element.

\begin{figure*}[t!]
%\centerline{\includegraphics{fig1.png}}
\centering
\includegraphics[width=1\textwidth]{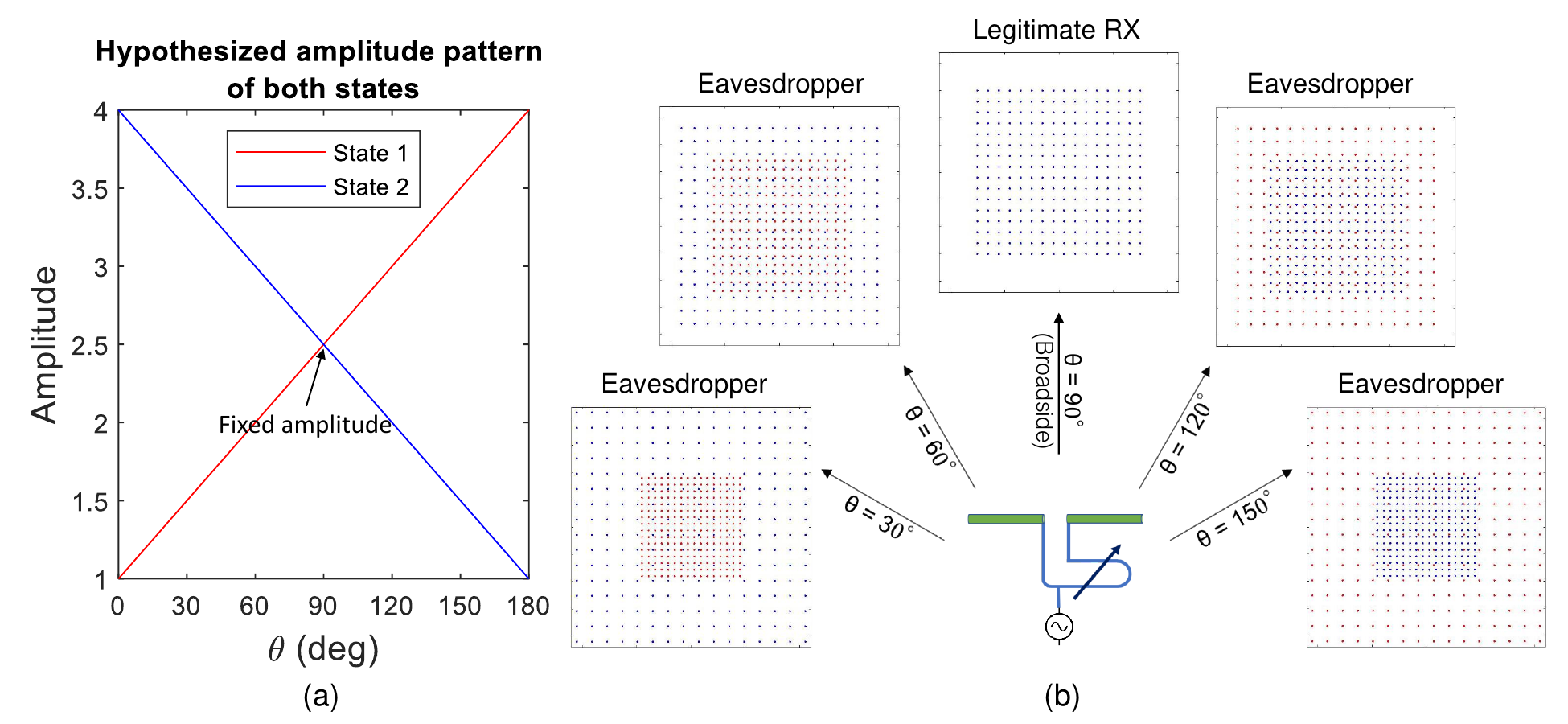}
\caption{(a) The proposed single-antenna directional modulation technique is based on rapidly switching between two asymmetric but mirrored states of the antenna, where one angle maintains a constant amplitude value between the two states. (b) By imparting a perturbation in the balun feeding a differentially-fed dipole antenna, asymmetric currents are excited on the arms of the dipole, generating an asymmetric pattern. By rapidly switching the balun perturbation between the two feeds using a switching structure, the antenna switches between the two states at a rate sufficiently fast to distort any transmitted or received data at all angles except where the pattern remains static, thereby achieve directional modulation at the physical layer.}
%Hypothesized Directional Modulation technique by constantly and rapidly changing an unbalanced feeding structure between two states to generate static field at the legitimate RX direction, and dynamics fields elsewhere. The hypothesized radiation pattern versus direction of the two states are purely amplitude (no phase) variations as shown in (a) and the corresponding constellations are plotted using same color code of each state (b). The constellations were generated by simulation using 300000 PRBS of 256-QAM modulation assuming a noise-free channel and uniform switching between the two states. All these constellations have same fixed axes limit.}
\label{Hypothesized_figure}
\end{figure*}

In this paper we demonstrate directional modulation in a single-port dipole antenna based on the use of a dynamic balun feeding structure. The balun actively switches between two states that  imbalance the output of the balun, generating small common-mode currents~\cite{iizuka1962effect}. The presence of common mode currents causes an asymmetric current distribution to manifest on the arms of the dipole. By switching the balun between two mirrored, imbalanced output states, the current distribution is likewise switched between two mirrored current states, generating radiation patterns with asymmetry. We rapidly switch the balun between the two states, creating a dynamic radiation pattern that is constant in only one intended direction and dynamic elsewhere, generating the directional modulation necessary for secure wireless transmission. Furthermore, as we discussed in~\cite{arisheh2023design}, the direction of the uncorrupted information may be steered. 
Compared to other directional modulation approaches, this work is the first to our knowledge that is implemented in a single, one-port antenna that has no switches or multiple feedpoints on the antenna. In particular, control of the differential and common mode currents that govern the dynamic radiation patterns manifests entirely in the balun, and thus can be implemented without changes to the antenna itself. As a result, the approach is simpler and more efficient than other directional modulation approaches.
The rest of the paper is organized as follows: In Section~\ref{Inv} we explore the inverse problem of determining the appropriate current distributions to generate the desired radiation patterns, then designing an antenna and balun structure to excite these currents. In Section~\ref{ant_design} we discuss the fabricated antenna and its measured radiation patterns. In Section~\ref{sec.measurements} the antenna is experimentally tested using a real-time wireless communications system. 
\section{Directional Modulation via Imbalanced Dipole Currents}
\label{Inv}

We consider the design of the dipole antenna by starting with the spatial and temporal form of a dynamic radiation pattern that exhibits sufficient relative differences at a wide range of angles outside of the desired information beam. Specifically, we look at the case of two mirrored states that are identical when reflected about the broadside direction, and that the dynamic antenna system will switch between rapidly. From these two mirrored radiation pattern states we determine appropriate current distributions on the arms of the dipole antenna. From this point, we determine a feeding structure that imparts the necessary currents.

%\subsection{Amplitude-Based Dynamic Radiation Patterns for Directional Modulation}
\label{RP}

%300000 "bits" were used (I know this for sure). Thus divide by 8 to know the number of symbols.

%The basic principle that we rely on to achieve a simple way for directional modulation by a single antenna was extensively discussed by theory, simulation, and measurements in our recent paper \cite{arisheh2023design}. That is, it relied on feeding a dipole from two feeding points, one at a time, controlled by a single external switch, without changing structure of the antenna. However, in this work we aim to achieve same principle by an effectively single port single antenna such that the antenna is fed from the exact single location. 

\begin{figure*}[t!]
    \centering
    \subfigure[]
    {
        \includegraphics[width=0.483\textwidth]{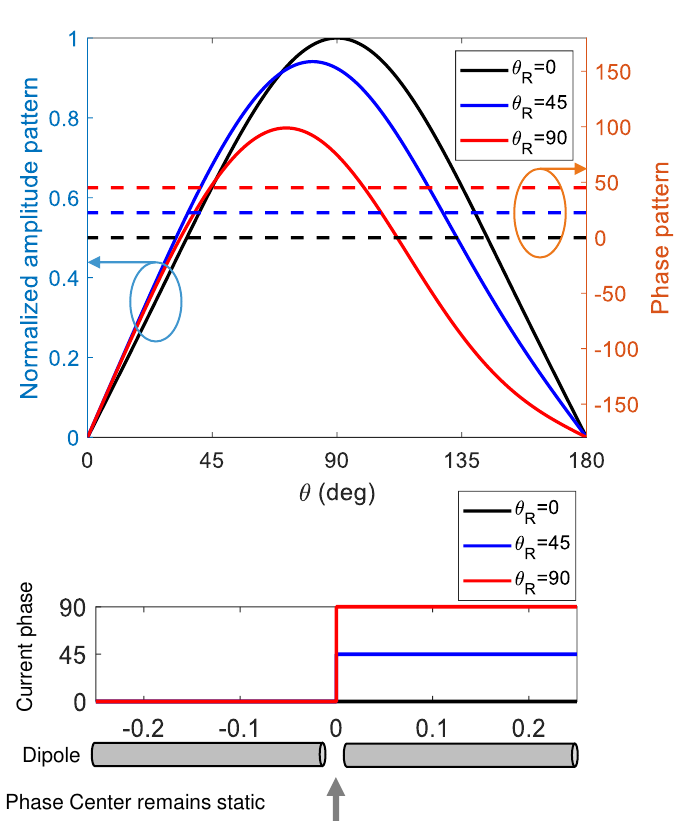}
        \label{R_steering}
    }
    \subfigure[]
    {
        \includegraphics[width=0.483\textwidth]{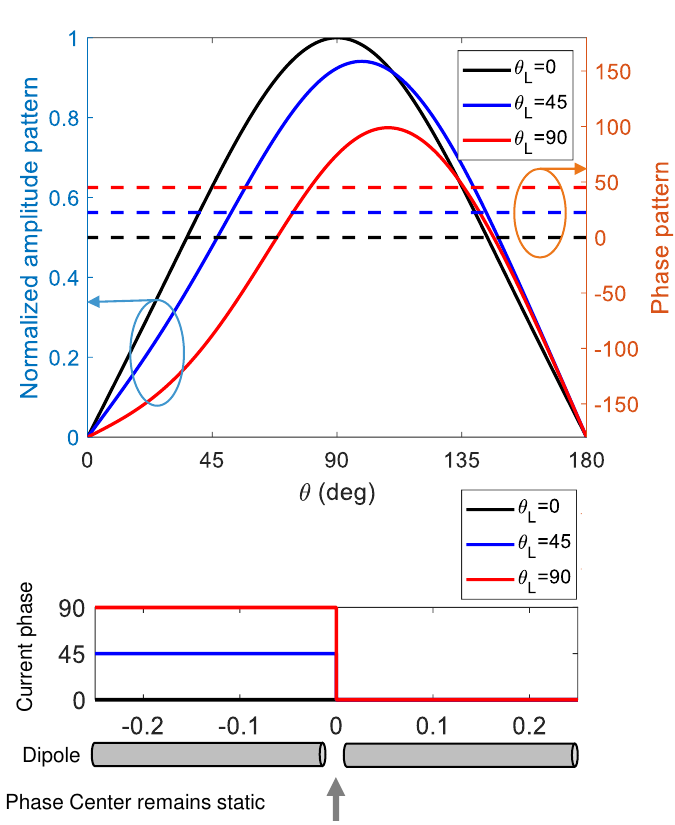}
        \label{L_steering}
    }
     \caption{Illustration of behavior of complex radiation pattern when imposing asymmetry to the dipole current by changing from a pure differential mode to a mixed mode with differential and common mode currents. This is done by imposing a phase lead of one of the dipole arms current relative to the other, however note that the phase center remains constant, as the phase pattern does not change between the two states. 
		%Mirrored cases are shown for the current which generate mirrored cases of the complex radiation pattern. 
		(a) Dipole current exhibiting asymmetry owing to common mode due to current lead of right arm $\theta_R$ compared to left arm. Three cases are shown: $\theta_R=0\degree,45\degree,90\degree$. The case $\theta_R=0$ represents the pure differential mode case which generates the standard radiation pattern. However, the more relative phase difference, the more steering and attenuation of amplitude pattern to left. However, phase pattern shape is same except for a jump equivalent to $\theta_R/2$. (b) Dipole current exhibiting asymmetry owing to common mode due to current lead of left arm $\theta_L$ compared to right arm.}
    \label{steering}
\end{figure*}

The general concept of implementing directional modulation in a single antenna is shown in Fig.~\ref{Hypothesized_figure}. 
%The general theoretical objective is summarized in Fig. \ref{Hypothesized_figure} which shows a single dipole antenna exhibiting directional modulation by dynamic fields. 
We hypothesize a state of current distribution on the dipole which generates a linear amplitude pattern as a function of angle as shown by the red trace in Fig.~\ref{Hypothesized_figure}(a). We assume that the antenna has a constant phase pattern. 
%versus direction (no phase variation) as shown in the red line in Fig. \ref{Hypothesized_figure}(a) while maintaining an isotropic radiation pattern. 
This defines State 1 of the antenna which exhibits the red constellations shown in Fig.~\ref{Hypothesized_figure}(b). 
Note that while the amplitude of the symbols change as a function of angle (and therefore the SNR will change), the overall structure of the constellation remains unchanged, allowing demodulation of the information from any angle.
Similarly, if we switch the dipole to State 2, defined by an inversely linear amplitude pattern versus angle, but with opposite slope to State 1, as shown in Fig.~\ref{Hypothesized_figure}(a), then we obtain the blue constellations as shown in Fig.~\ref{Hypothesized_figure}(b), which again allows demodulation of the data from any angle.
Now, if the antenna is rapidly switched between States 1 and 2, the resultant constellations will be a combination of the red and blue constellations. At the point where the amplitude between the two states is equal (here at broadside, $\theta=90\degree$, although this can be placed at angle single or multiple angles depending on the radiation pattern), the constellations are identical, and thus the information remains unchanged and can be directly demodulated.
At all other angles, the superposition of the two constellations results in distorted constellations, making it more difficult to demodulate the information and thereby creating a layer of directionally-dependent security. We refer to the angles wherein the data is unmodulated by the dynamics of the switched antenna pattern as the \textit{information beam}. Further details on the general concept are given in~\cite{arisheh2023design}.  

%If we keep antenna working in State 1 then constellations in all directions will be standard thus allowing both legitimate and eavesdropping receivers to make standard detection. Similarly, if we switch the dipole to State 2 which generates an inversely linear amplitude pattern versus direction (no phase variation) as shown in Fig. \ref{Hypothesized_figure}(a), then we obtain the blue constellations as shown in Fig . \ref{Hypothesized_figure}(b), which again are standard and thus allowing any RX for standard detection from whichever direction. However, if we perform a rapidly and constantly random switching, then the legitimate RX will still maintain standard constellation due to static amplitude in his direction; however, simultaneously, other directions will receive distorted constellations owing to dynamic amplitudes there. Thus, the more offset from broadside, the more dynamic fields, and more distortion making it hard for eavesdroppers to perform detection. This behaviour allows information to come through only in the broadside direction (information beam), while being distorted elsewhere. This information beam can be simply steered by moving the fixed phase point from broadside. Interestingly, the amplitude dynamics works only in large QAM modulation orders, and the larger causes narrower information beamwidth. More elaboration on the technique and its details in our previous work \cite{arisheh2023design}.  

The complementary case of using asymmetric phase patterns in each state is also viable (see~\cite{arisheh2023design}). In this work, we focus on the use of amplitude-only dynamics. Ultimately, a combination of the two dynamics may yield improved results; however in this work we focus only on amplitude.
%However, using phase-only dynamics yields high SNR at all angles, whereas 

%The complementary case of relying on phase pattern dynamics while having isotropic phase pattern was discussed also \cite{arisheh2023design}. \textcolor{blue}{However, amplitude dynamics come with price of losing gain in eavesdroppers directions as it is always desired to reduce power in their direction. This also costs power loss when trying to steer information beam. Thus, the ideal solution is by relying on phase pattern dynamics while having amplitude pattern static or quasi-static to minimize amount of losses.}

%Such radiation patterns characteristics were reported in literature to achieve Directional Modulation by arrays or multiport antennas. However, they have never been reported via a single port single antenna.

\subsection{Generating Asymmetric Dipole Radiation Patterns Via Asymmetric Current Excitation}
\label{current_stage}

We choose a dipole antenna as an example to study the generation of asymmetric amplitude patterns. Dipole antennas are relatively simple antenna structures, and have current distributions that can be characterized to first order in a fairly straight-forward mathematical formulation. From the current distributions the far-field radiation patterns can be calculated. The dipole antenna also has two arms on which the currents can be excited differently, providing current asymmetry, and thus provides a direct analysis of the radiation patterns from simple changes to the currents on the antenna arms.

The current distribution on a standard half-wavelength dipole antenna exhibits a sinusoidal shape that goes to zero at the ends and can be given to first order by~\cite{king1956theory}
\begin{equation}
I(z) = I_m \sin{k\left(\frac{L}{2}-|z|\right)}
\end{equation}
where $k = \frac{2\pi}{\lambda}$ with $\lambda$ the wavelength, $I_m$ is the maximum current amplitude, and $L$ is the total length of the dipole antenna. 
We can consider the current as the superposition of the currents on the two arms of the dipole antenna by
\begin{equation}
I(z) = e^{j\theta_L}I_L(z) + e^{j\theta_R}I_R(z)
\end{equation}
where $I_L$ and $I_R$ are currents on left and right arms, respectively, and $\theta_L$ and $\theta_R$ are relative phases for current on left and right arms, respectively.
From this formulation, we can study the impact of relative differences between the currents on the two arms of the antenna on the subsequent radiation patterns. We will consider the case where each complex current has a relative phase shift.
Note that when $\theta_L = \theta_R$ the currents are symmetric, and result in a purely differential current input to the dipole, which is the traditional feeding approach. When $\theta_L \neq \theta_R$, however, asymmetric currents manifest that yields both differential and common mode currents. As we show later, by implementing currents that are largely differential but have a common mode component, sufficient asymmetry in the radiation patterns can be generated to create the desired asymmetric radiation patterns with minimal impact on the radiated power.

The magnetic vector potential in the far field can be calculated from the currents by
\begin{equation}
\textbf{A}(\textbf{r}) \approx \hat{z} \frac{e^{-jkr}}{4\pi r} \int_{-L/2}^{L/2} I(z') e^{jk{z'}\mathrm{\cos{{\mathrm{\theta}}}}} d{z'} 
\end{equation}
where the primed coordinates refer to the source location and unprimed coordinates to the observation point.
From this, the far-field magnetic field intensity is 

\begin{align}
{\mathbf{H}} &= \nabla \times \mathbf{{A}} \\ 
&\approx \hat{\phi}jk\frac{e^{-jkr}}{4\pi r} \sin{\theta} \int_{-L/2}^{L/2} I(z') e^{jk{z'}\cos{\theta}}d{z'} \\ 
&\approx \hat{\phi}jk \frac{e^{-jkr}}{4\pi r} \sin{\theta} \Biggl\{ \int_{-L/2}^{0} e^{j\theta_L} I_L(z') e^{jk{z'}\cos{\theta}} d{z'} + \\ & \; \; \; \; \; \; \; \; \; \; \, \; \; \; \; \; \; \; \; \; \; \; \; \; \; \; \; \: \; \; \;  \int_{0}^{L/2} e^{j\theta_R} I_R(z') \: e^{jk{z'}\cos{\theta}} \,d{z'} \Biggl\} \\ 
&= {\mathbf{H_L}} + {\mathbf{H_R}}
\label{Eq_4}
\end{align}

from which the far-field electric field intensity can be obtained.
The far-field patterns can be seen to be approximately related to the current densities through a Fourier transform. We'll consider the magnitude of the Fourier transform of the currents, which is thus proportional to the amplitude pattern of the antenna.

In Fig. \ref{R_steering}, we explore a few examples of the magnitude of the radiation pattern as a result of different phase excitations
The case of a standard (purely differential mode) dipole current ($\theta_R=\theta_L=0$) is shown in the black curve. With a purely differential current feeding the antenna, the currents on the antenna arms are always in the same direction and of equal amplitude at points symmetric about the feed. Thus, the pattern in both amplitude and phase is also symmetric. The phase pattern is also constant with angle. 

When $\theta_R \neq \theta_L$, common mode currents manifest, which run in opposite directions on the antenna arms. The superposition of the differential mode and common mode currents results in asymmetries on the currents, as shown by the plots of the antenna current phase. When the relative phase between the two currents is $45\degree$ in either the left or right arm, the magnitude of the far field radiation pattern exhibits a small asymmetry, however the phase pattern, while undergoing a bulk phase change, remains constant in angle. It can be seen that implementing the phase shift on either the left or right arms yields mirrored amplitude states, which generates the desired differential amplitude patterns as described above: at broadside ($\theta = 90\degree$) the amplitude pattern is the same between the two states, while away from broadside each state yields a different amplitude value. Increasing the relative phase to $90\degree$ results in greater asymmetry and greater differences between the two states, however at the cost of reduced amplitude in the radiation pattern. The reduced amplitude is a consequence of having larger common mode currents on the antenna.

\subsection{Excitation of Asymmetric Dipole Currents}

\begin{figure}[t!]
%\centerline{\includegraphics{fig1.png}}
\centering
\subfigure[]
    {
\includegraphics[width=0.4\textwidth]{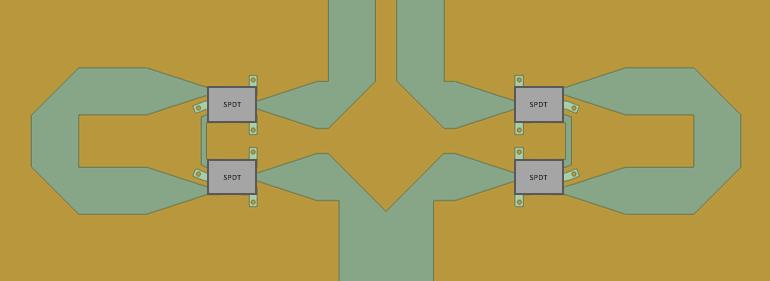}

}
\subfigure[]
    {

\includegraphics[width=0.4\textwidth]{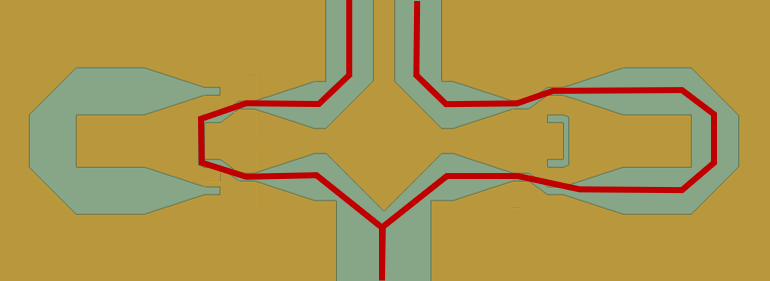}

}
\subfigure[]
    {
\includegraphics[width=0.4\textwidth]{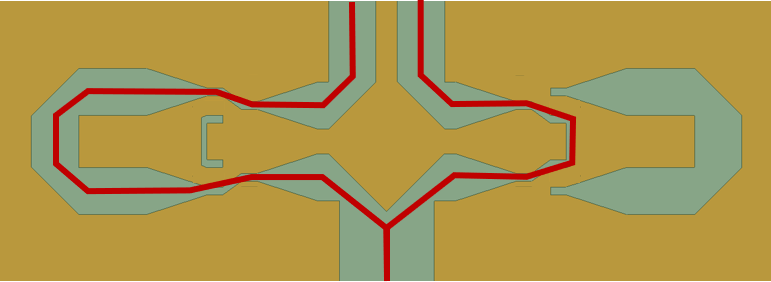}

}

\caption{(a) Design of the dynamic balun with four single-pole double-throw (SPDT) switches to change the current paths. (b) Current path in State 1. (c) Current path in State 2.}
\label{balun}
\end{figure}

As described above, the concept is dependent on feeding the arms of the dipole antenna with asymmetric currents. To accomplish this, we designed a dynamic balun based on a traditional microstrip balun design and switches that are constantly switching the current paths between two mirrored states, each of which provides an asymmetric current feed to the dipole arms. Fig.~\ref{balun}(a) shows the general design. The balun is designed to impart an imbalance between the two lines in two mirrored states. The switches are used to rapidly change the feeding currents between two states, as shown in Figs.~\ref{balun}(b)-(c). In either state, the input signal is converted to a differential signal, however the lengths are  imbalanced from ideal (see dimensions in the following section) to impart the desired shift in the amplitude pattern. Switching between the two states switches the radiation pattern between the two mirrored states as illustrated in Fig.~\ref{steering}. By rapidly switching between the two states, the desired dynamic radiation pattern is obtained.

\begin{figure}[t!]
%\centerline{\includegraphics{fig1.png}}
\centering
\includegraphics[width=0.49\textwidth]{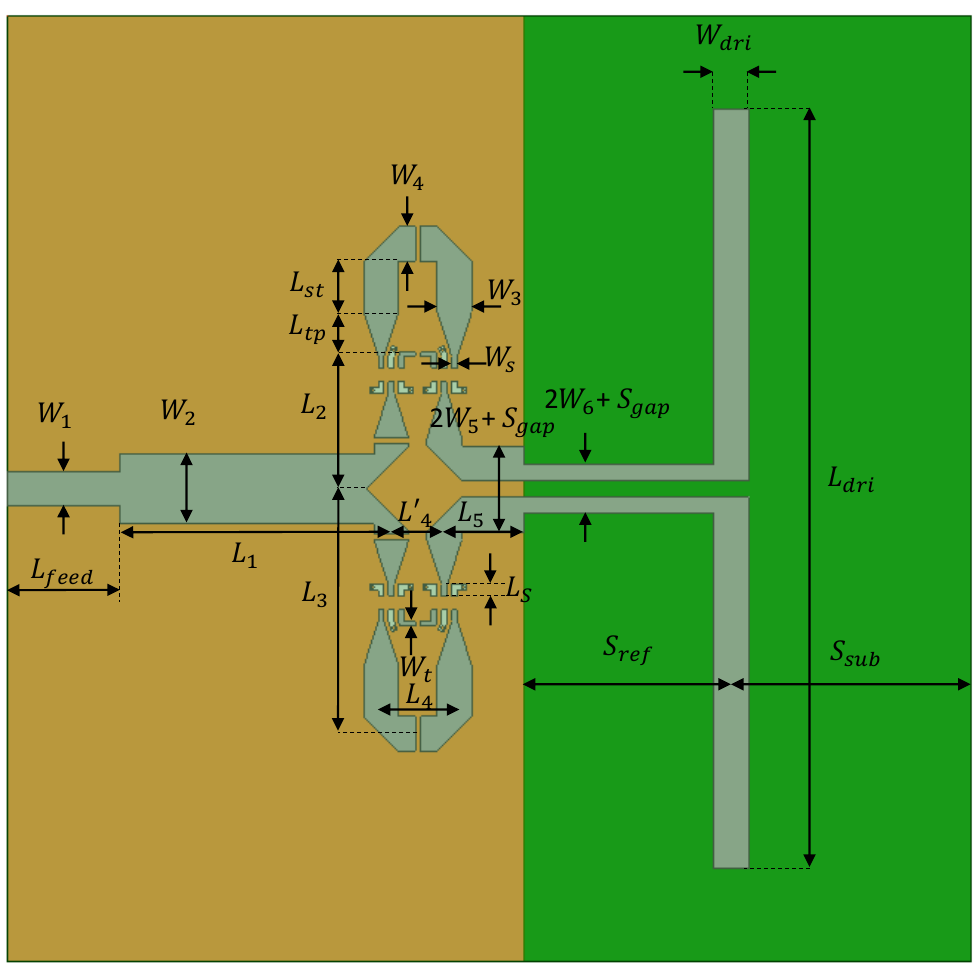}
\caption{Design layout of the dipole antenna and dynamic balun.}
\label{Dims}
\end{figure}

%\textcolor{green}{Trade-off between losses and dynamics. Dunno if we should put common mode content increase due to steering and its accompanies losses}

%\textcolor{green}{Put the figure that I show external and internal control of the dipole and that frequency affects the internal. The external is related to information beam steering (RP scaling/ phase pattern jump).}

%\textcolor{green}{(Maybe I can say, ON OFF switching by FT naturally give phase pattern while here only some amount of current goes between elements similar to Jacob).}

%\textcolor{blue}{I did not talk about the possibility of monopole radiation to contribute since I think it should radiate when there is unbalance. We also did not say that the ground will act as a reflector and thus this is Quasi-Uda I think not a pure dipole.}

%\textcolor{green}{WARNING: Here losses due to the null without the feeding part. Feeding losses (and contribution to PC movement) are not included in this analysis. i.e. Feeding part is not included. I think I can imagine a fictitious feeding and just find how much input impedance affected by common mode and thus find S11 loss!}

\section{Antenna Fabrication and Measurement}
\label{ant_design}

%Put two states of the designed antenna and show current, etc. For steering I might need a vector plot.

%I can touch on common/balanced analysis and I can put my antenna and make the balun balanced to show that the phase pattern became symmetric and then show my current case.

%Show by simulation Phase Center Ideal Motion Between Two Points.

%I can show that I have amplitude dynamics with small phase dynamics at 1.49 GHz.

%I can talk about reconfigurable antennas, if used for DM then lose a lot of power for secure beam steering.

%\subsection{We can go to upper frequencies to possible get different PC (i.e. different phase pattern dynamics) and/or different amplitude dynamics. Generally, ignoring gain, any frequency can give good dynamics in case either amplitude or phase is a little asymmetric.}

%Quasi-Yagi antennas have been used in wide range of applications \cite{kasabegoudar2022quasi}. In \cite{qin2013beam}, it is being used for beam switching pattern reconfigurability by controlling the balun length using PIN diodes; however, here the application is for a dynamically switched antenna for DM. 

\begin{figure}[t!]
    \centering
    %\subfigure[]
    %{
        \includegraphics[width=0.175\textwidth]{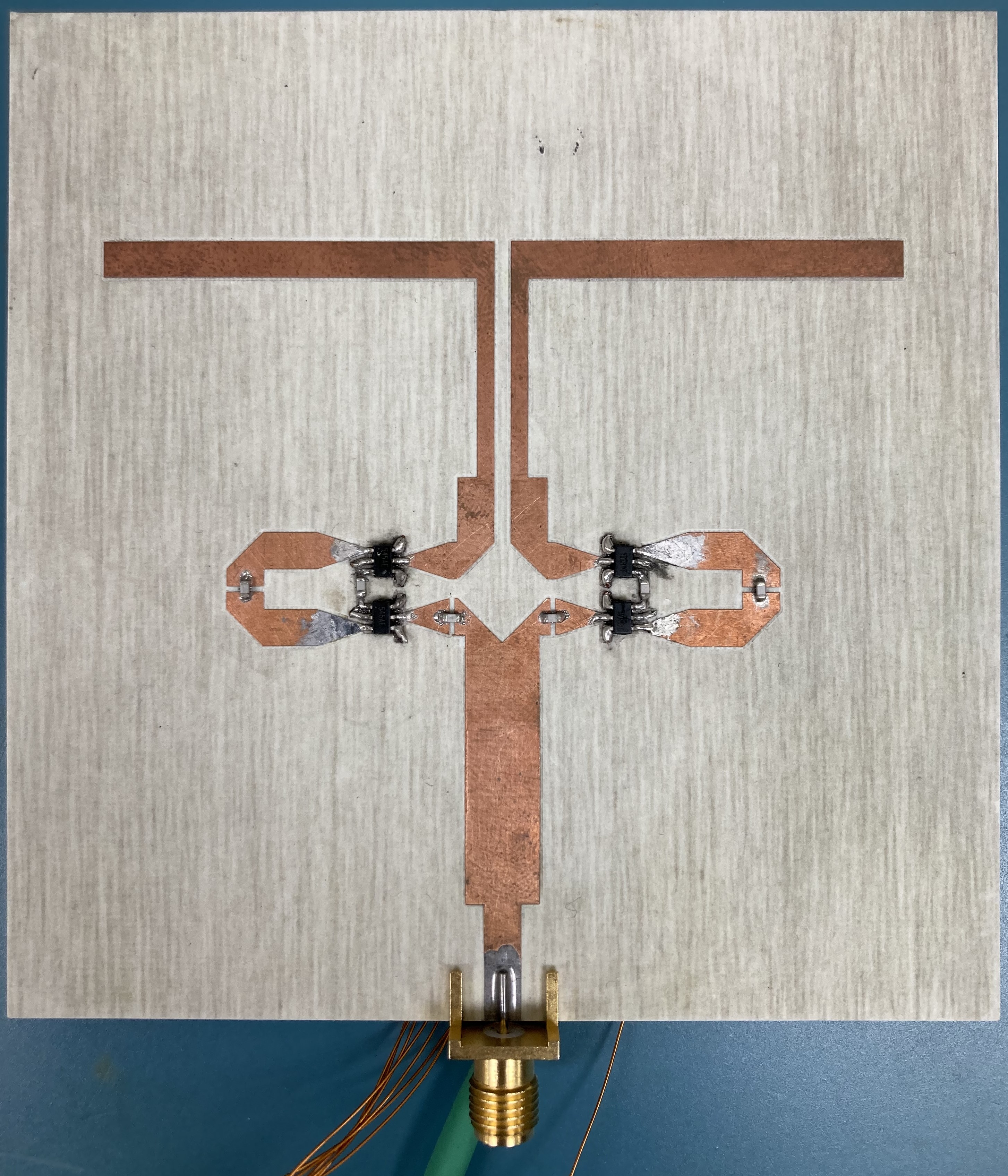}
				\includegraphics[width=0.205\textwidth, angle=90]{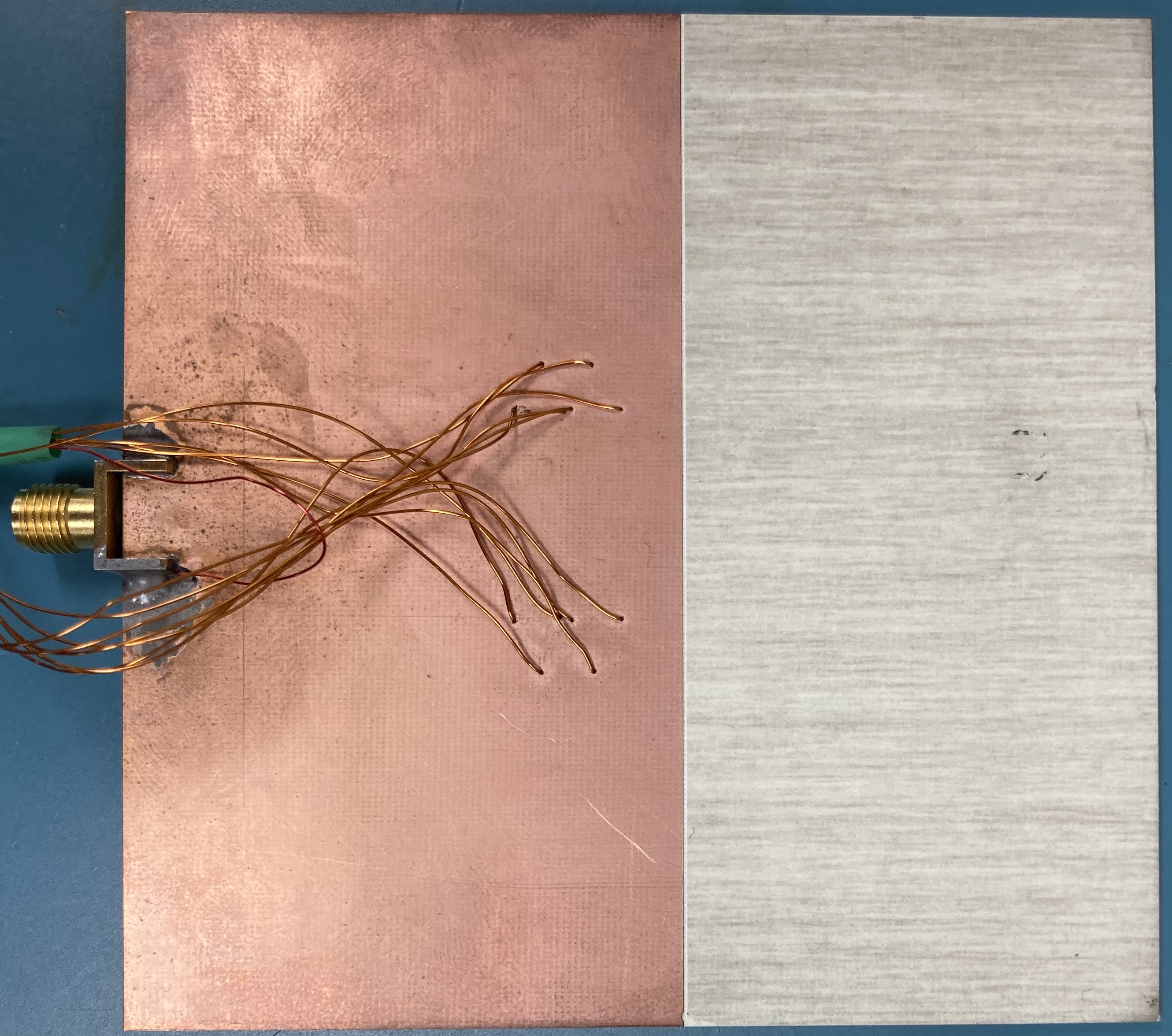}
       %\label{front}
    %}
   
    %\subfigure[]
    %{
        
        %\label{back}
    %}
    \caption{(Left) Front and (Right) back of the fabricated active antenna. The balun uses four surface-mount SPDT switches and six capacitors.}
\label{Antenna_pic}
\end{figure}

The antenna and dynamic balun are based on a traditional differential fed dipole antenna (e.g.~\cite{kaneda1999broad,deal2000new}) and a balun structure with switched to accommodate two different path lengths to the differential feed. The design and dimensions are shown in Fig.~\ref{Dims} and Table~\ref{table}. The substrate was Rogers 4350B with 1.524 mm thickness and the ground plane was partially extended as shown in brown in Fig.~\ref{Dims}. 
The fabricated dynamic antenna was configured using four switches (HMC545AE) and six dc-blocking capacitors (0603ZA101JAT2A) as shown in Fig.~\ref{Antenna_pic}. The switches were connected to a microcontroller (MCU) that configured the switches to create a static State 1, a static State 2, or dynamic switching between the two states according to an input square wave signal. 
%The convention of State 1 or State 2 is illustrated in Fig.~\ref{Convention}.

\begin{table}[]
\centering
\caption{Antenna Dimensions}
\label{table}
\begin{tabular}{|*{12}{p{2.9cm}|}}
\hline
Parameter       & Value \\ \hline
$W_1,W_3,W_4,W_5,W_{dri}$     & $3.285$ mm  \\ \hline
$W_2$   & $6.57$ mm  \\ \hline
$W_t$        & $\approx0.39$ mm        \\ \hline
$W_S$       & $0.55$ mm      \\ \hline
$W_6,S_{gap}$       & $1.5$ mm      \\ \hline
$L_{feed}$       & $10.57$ mm     \\ \hline
$L_{1}$         & $\approx25.5$ mm      \\ \hline
$L_{2}$         & $\approx12.68$ mm      \\ \hline
$L_{3}$         & $\approx 23.04$ mm      \\ \hline
$L_{4}$         & $6.9$ mm      \\ \hline
$L_{4}^{'}$         & $5$ mm      \\ \hline
$L_{tp}$         & $4.2$ mm      \\ \hline
$L_{st}$         & $4.715$ mm      \\ \hline
$L_{dri}$         & $71.5$ mm      \\ \hline
$S_{ref}$         & $19.5$ mm      \\ \hline
$S_{sub}$       & $22.5$ mm      \\ \hline
\end{tabular}
\end{table}

%\begin{figure}[]
%%\centerline{\includegraphics{fig1.png}}
%\centering
%\includegraphics[width=0.48\textwidth]{Images/Convention.jpg}
%\caption{The balun part of the fabricated active antenna showing four SPDT switches and 6 capacitors in addition to convention of antenna State 1 and State 2.}
%\label{Convention}
%\end{figure}

\begin{figure}[]
%\centerline{\includegraphics{fig1.png}}
\centering
\includegraphics[width=0.4\textwidth]{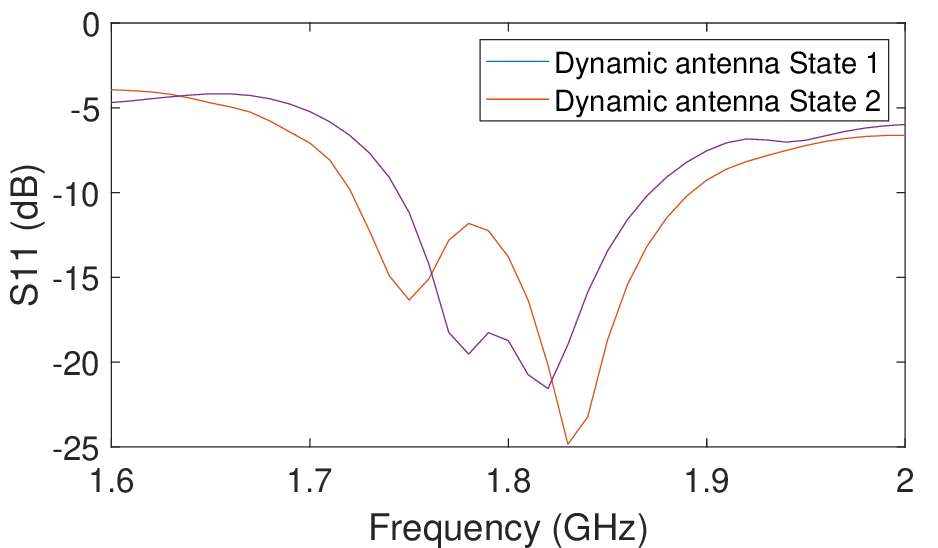}
\caption{S11 versus frequency for both states of the fabricated active antenna.}
\label{S11}
\end{figure}

\begin{figure}[t]
%\centerline{\includegraphics{fig1.png}}
\centering
\includegraphics[width=0.48\textwidth]{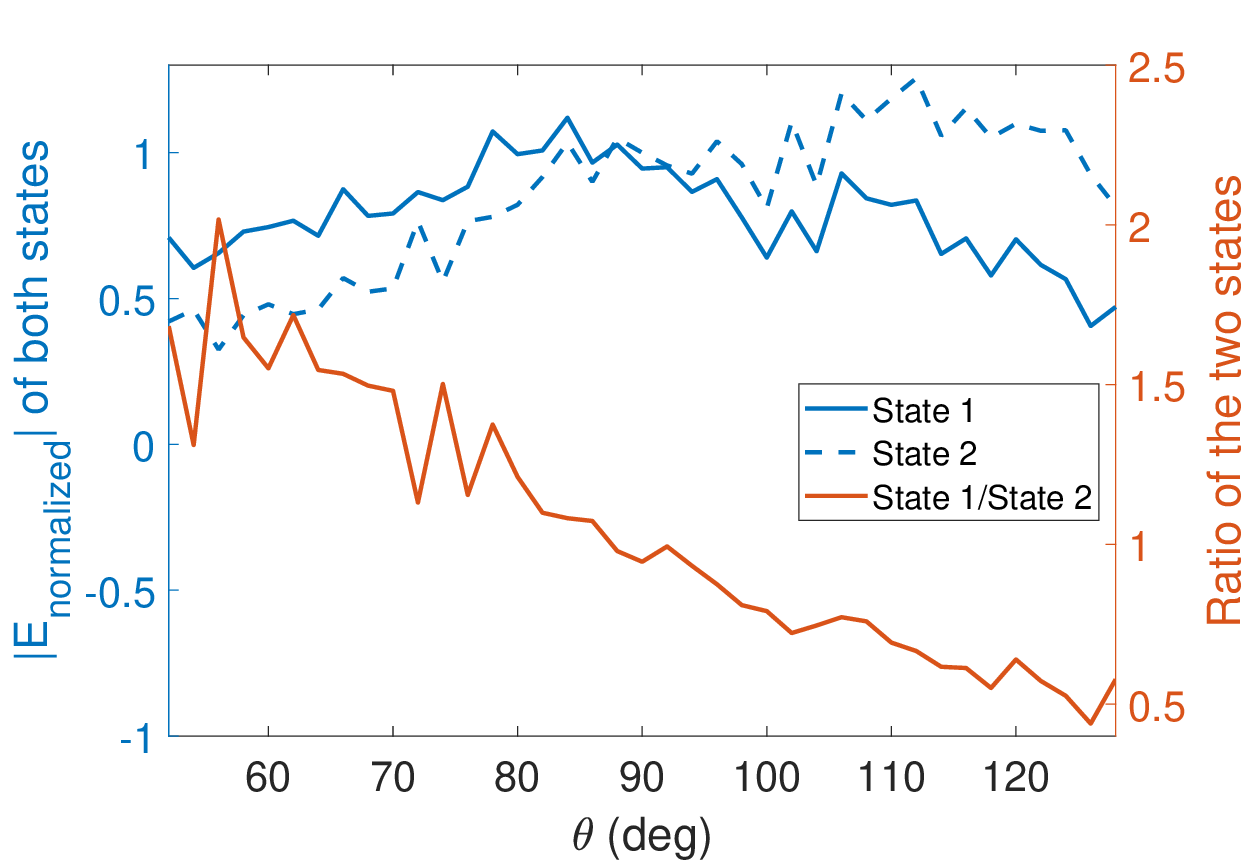}
\caption{Amplitude patterns of both states at 1.86 GHz and the ratio of the two states. When the ratio is near unity the modulation due to the antenna dynamics remains small, while when the ratio is $\neq 1$ the difference in the amplitude patterns induces additional modulation on the signal.}
\label{1.86_amplitude}
\end{figure}

\begin{figure}[t]
%\centerline{\includegraphics{fig1.png}}
\centering
\includegraphics[width=0.48\textwidth]{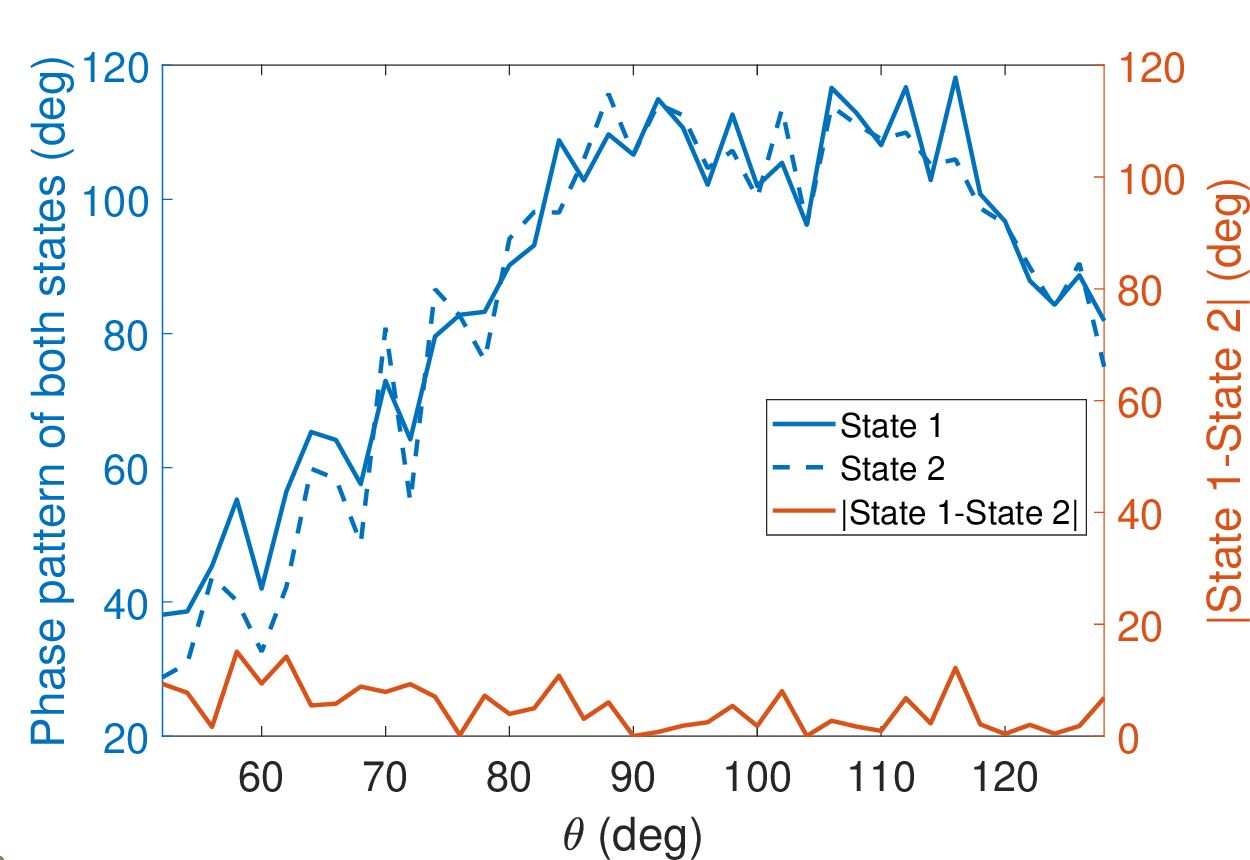}
\caption{Phase patterns of both states at 1.86 GHz and the difference between the two states. Since the differential phase is small at all angles, the directional modulation is due only to the amplitude patterns.}
\label{1.86_phase}
\end{figure}

%The measured return loss of both states exhibits resonance at more than one region as shown in Fig. \ref{S11}. At 1.86 GHz, both states have S11 $<$ -10 dB, and the gain is approximated to be around +1.5 dBi at the center of the end-fire direction. However, our previous antenna gain had been estimated to be around -4 dBi \cite{arisheh2023design}. The input impedance of each switch is 50 ohm, while the characteristic impedance of any line that is directly soldered to any switch leg is 116.4 ohm at 1.86 GHz (since $W_S=0.55mm$). This results in a reflection coefficient of around 0.39. 

The S11 of the antenna in both states is plotted in Fig.~\ref{S11}, showing a resonance frequency centered around 1.8 GHz. At this frequency the gain was 1.5 dBi in both states.
The major beam amplitude and phase patterns and their ratio at 1.86 GHz are illustrated in Fig.~\ref{1.86_amplitude} and Fig.~\ref{1.86_phase}, respectively. 
%The major beam is assumed to be $\pm 38\degree$ from the center and is measured with $2\degree$ steps. 
The ratio of amplitude patterns shown in Fig. \ref{1.86_amplitude} exhibits a value of more than 1.14 for most angles outside the region between $82\degree$ and $96 \degree$. 
Referring to the theoretical analysis in our previous work~\cite{arisheh2023design}, we see that outside of the $82-96\degree$ window this is sufficient to result in errors for 256-QAM even in a noise-free environment and without any phase dynamics. Thus, this amplitude pattern generally satisfies the conditions for generating directional modulation since the amplitude is almost static at the center, and there is sufficient dynamics at off-side directions to distort large order QAM signals. 
Also, the power difference between the radiation patterns of the two states is minimal for many angles, therefore the information beam could be steered within a region around the central direction without significant loss. Thus, the amplitude pattern at 1.86 GHz exhibits dynamics that are large enough for directional modulation and small enough to support information beam steering. 
The phase patterns between the two states differ minimally, generating the desired amplitude-only dynamics.
%Also, the phase pattern (after adding a constant shift to calibrate it at the center) shows small phase dynamics for some directions as shown in Fig. \ref{1.86_phase}, which will aid to further increase the errors at these directions, especially in directions having small or negligible amplitude dynamics, which will increase the directional modulation. 
The fact that the phase dynamics are generally small means that the phase center of the antennas is quasi-static and that currents on the dipole differ mainly in phase, not in magnitude as explained in Section.~\ref{current_stage} and Fig.~\ref{steering}. 
%By comparing Fig.~\ref{steering} and Fig.~\ref{1.86_amplitude}, we conclude that phase difference between the two dipole currents can be generally $\le 90\degree$. 
Note that the phase patterns exhibited a constant differential phase bias that is normalized in Fig.~\ref{1.86_phase}. This constant bias is due to different phase delays in the switch components, and could be addressed in various ways, such as by characterizing the phase delays of the switches or by choosing devices with minimal phase delay. In this work, we calibrate the phase bias using an external calibration tuned to the switching frequency of the antenna (described further in the following section). While this approach requires some additional circuitry, it serves effectively as a proof-of-concept to demonstrate the proposed antenna.

\section{Experimental Evaluation}
\label{sec.measurements}

\begin{figure}[t!]
%\centerline{\includegraphics{fig1.png}}
\centering
\includegraphics[width=0.48\textwidth]{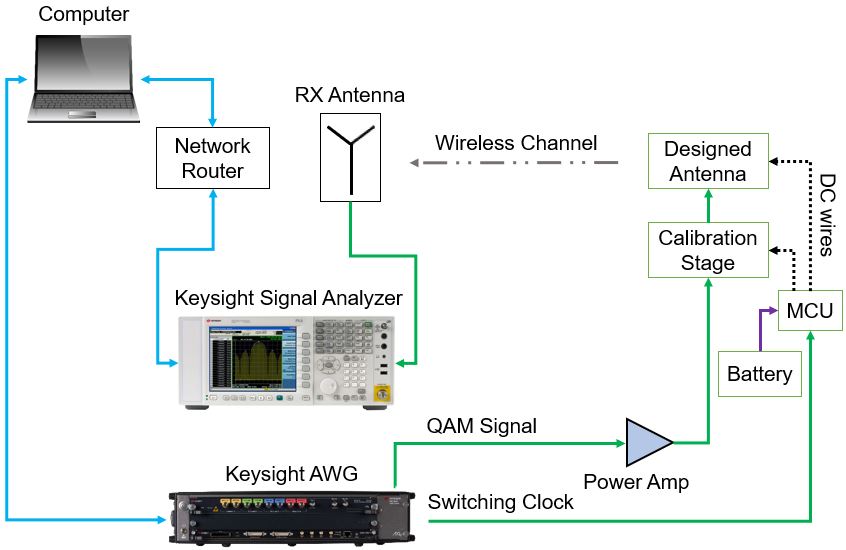}
\caption{Block diagram of the real-time communications experiment.}
\label{Setup}
\end{figure}

%\begin{figure}[]
%%\centerline{\includegraphics{fig1.png}}
%\centering
%\includegraphics[width=0.48\textwidth]{Images/Calibration.JPG}
%\caption{A dynamic calibration stage to make a constant shift of amplitude and phase of one of the states compared to the other. However, it does not affect the shape of the radiation pattern versus direction.}
%\label{Calibration}
%\end{figure}

\begin{figure}[t!]
    \centering
    \subfigure[]
    {
        \includegraphics[width=0.48\textwidth]{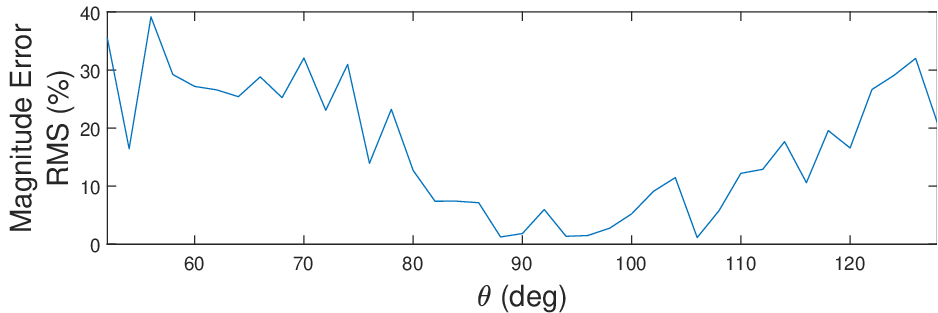}
       \label{Fig:Qa}
    }
    \\
    \subfigure[]
    {
        \includegraphics[width=0.48\textwidth]{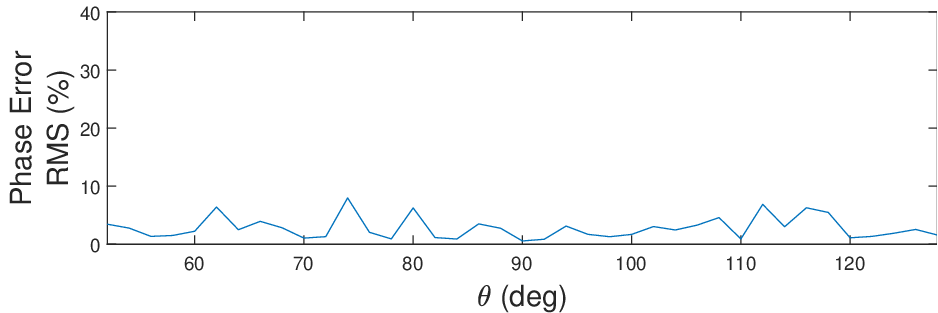}
        \label{Fig:Qb}
    }
    \\
    \subfigure[]
    {
        \includegraphics[width=0.48\textwidth]{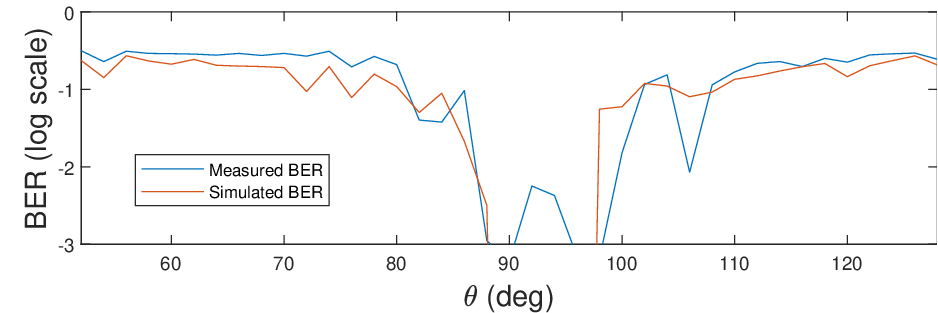}
        \label{Fig:Qc}
    }
    \caption{(a) Measured magnitude error of the received signals. The error is low near broadside and larger elsewhere as a result of switching between the two states. (b) Measured phase error showing low errors at all angles. (c) Measured and simulated BER, demonstrating low errors in an $8\degree$ window near broadside and high errors elsewhere, demonstrating the  directional modulation efficacy.}
    
\label{measurements}
\end{figure}

%The experimental evaluation for the proposed dynamic antenna to exhibit Directional Modulation is conducted by a live wireless communication system. 
We evaluated the functionality of the dynamic antenna system in a real-time communications system, the block diagram of which is shown in Fig.~\ref{Setup}. 
A Keysight M8190A Arbitrary Waveform Generator (AWG) is used to generate a quadrature-amplitude modulated (QAM) signal that is fed to the calibration stage which corrects the constant phase bias as discussed above, and then the signal is fed to the dynamic antenna. The dynamic antenna was switched at 3 kHz by a microcontroller (MCU) which was controlled by a clock signal generated by the AWG. The height of transmitter and receiver were 1.3 m and 1.5 m, respectively, and were separated by 3.4 m in a semi-anechoic environment. The signal was received by a horn antenna and demodulated by a Keysight N9030A PXA Signal Analyzer. A PC was connected to the PXA and used both IQtools and Vector Signal Analysis software to control the generation and analysis of the received signal.

\begin{figure*}[t!]
%\centerline{\includegraphics{fig1.png}}
\centering
\includegraphics[width=0.8\textwidth]{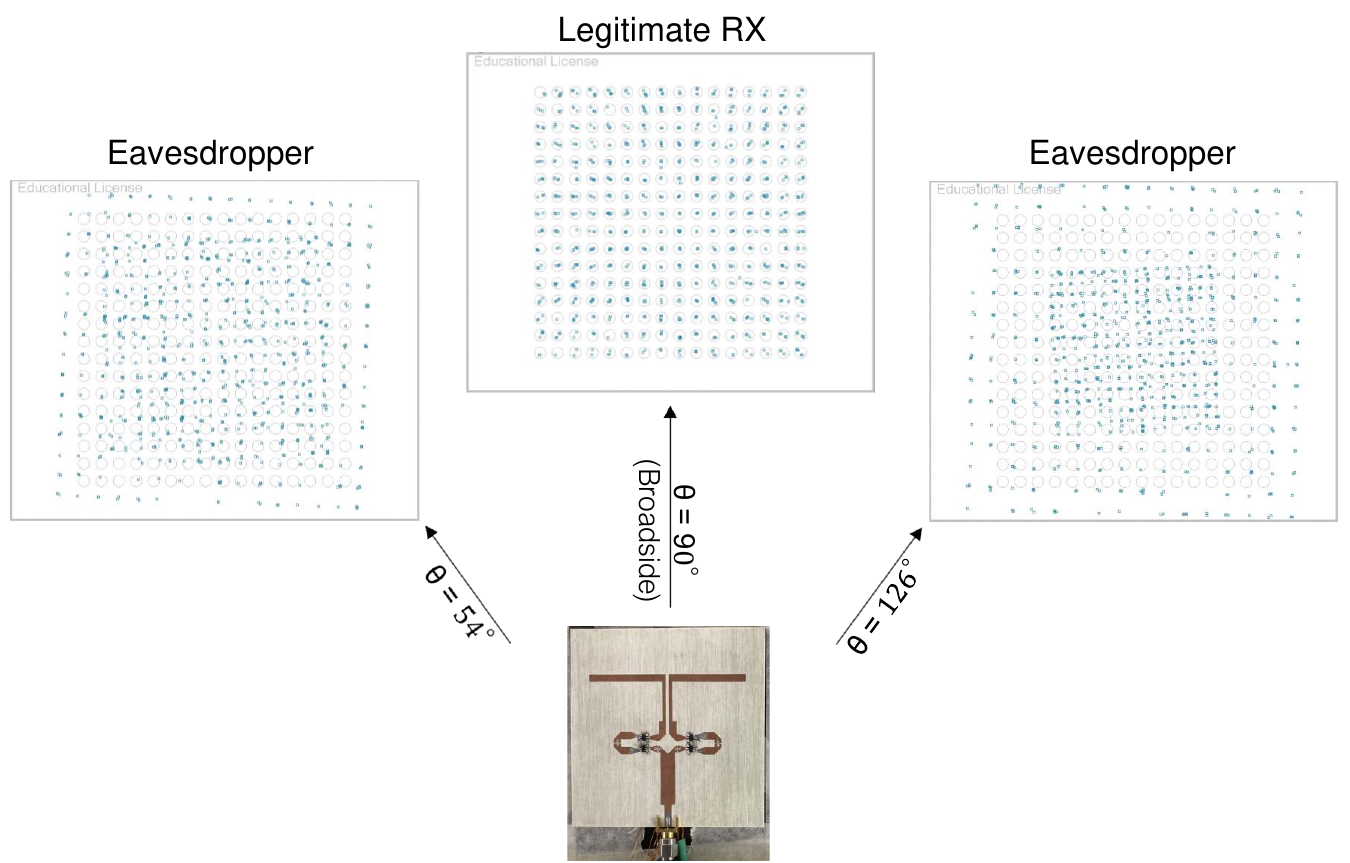}
\caption{Experimental demonstration of wireless security by a single port single antenna. The data constellations were measured using 900 256-QAM symbols. The constellation at broadside is standard, having small amplitude and phase errors, for recoverable data transmission. At other angles the impact of the differential amplitude states is evident, as the two superposition of the two constellations is visibly evident. By switching sufficiently fast, the data at these angles is unrecoverable by the receiver.}
%Usedmixedfiledon'tworry
\label{Constellations}
\end{figure*}

The calibration stage used to correct the static phase difference between the two states was implemented using two coaxial cables and two switches, with a variable phase shifter on one line that was adjusted to offset the static phase difference. The switches were controlled with the same 3 kHz signal as the antenna. Note that the static phase difference could also be corrected by including a phase shifter on one of the balun arms.

%Although our dynamic antenna has a single feeding port, a dynamic calibration stage was connected using two external SPDT switches (HMC595A) for calibration purposes of the two states as shown in Fig. \ref{Calibration}. These external switches, in addition to the four surface mount switches on the antenna, are controlled by a microcontroller to create the two states and switch between them. The calibration was used to make the complex radiated field static at the center of end-fire direction (Angle $90\degree$). The calibration is adjusted by a phase shifter, attenuators and different coaxial wires lengths. Thus, depending on the current states, the signal goes through either of the two branches of the calibration stage. Interestingly, this calibration stage can be achieved either using AWG signaling or by hardware. However, here we achieved it by a cascaded external hardware which makes it self-contained with the design which further prove that the technique can be used in a black box fashion (transparent to the system) without any need to change the input conventional signaling. In addition, calibration part can be also used to get static complex field at another angle and thus steer the information beam to that angle \cite{arisheh2023design}.

Measurements were conducted over angles between $52\degree$ and $128\degree$ in $2\degree$ steps. The transmitted communication signal was a 256-QAM signal with a 1 MS/s symbol rate at a center frequency of 1.86 GHz. Specifically, a $2^{11}-1$ PRBS signal with a length of 30 000 symbols was transmitted and were captured at each measured angle.
 %The bit error ratio (BER) and other error metrics were measured using 72 kbit and were lively viewed in the computer. Note that reception was done assuming that the receiver is aware of transmission such as  frequency, modulation order, and symbol rate which is considered as a bad scenario in terms of security.
The measured magnitude error, measured phase error, along with the measured and simulated bit error ratio (BER) are shown as a function of angle in Fig.~\ref{measurements}. 
The simulated BER included in Fig. \ref{Fig:Qc} was done in MATLAB using 72 kbit of gray-coded 256-QAM and uniform switching assuming a noise-free channel using the measured fields provided in Section. \ref{ant_design} after making a numeric calibration for both amplitude and phase in MATLAB at the central direction.
The magnitude error is the rms average of the difference in amplitude between the measured signal and the transmitted signal. The
phase error is the phase difference between the measured and transmitted averaged over all symbol points.
The magnitude error and phase error show small values at broadside ($90\degree$) as designed. The phase error is small at all angles, while the magnitude error increases at angles away from broadside, which also matches the design goals described above.
%Magnitude error is in general as expected from the radiation patterns in Fig. \ref{1.86_amplitude} since it is small in center while, generally, is larger elsewhere. Also, the phase error agrees with Fig. \ref{1.86_phase} since it is generally small and goes up/down without any monotonic behaviour; and thus the phase center was in general quasi-static. 
Subsequently, the measured BER is larger than $10^{-3}$ for all angles except within an $8\degree$ window at broadside (matching the simulated BER) where the information remains recoverable. 
%It shows a general agreement with the measured BER. %Since dynamics of the complex radiation pattern are gradual, then it is not unexpected to get some information transferred close to the calibrated region since dynamics will still be insufficient to create errors.
Measured data constellations at broadside and two off-side directions are shown in Fig.~\ref{Constellations}, display the desired dynamics as indicated in Fig.~\ref{Hypothesized_figure}.

 \section{Conclusion}
 \label{Conclusion}
 
A single port single antenna has been introduced to achieve physical layer security via directional modulation approach for wireless applications at 1.86 GHz with gain of 1.5 dBi. Prior similar works \cite{2017,2021,cabrera2022multibeam,zandamela2022directional,arisheh2023design} have all required multiple inputs or multiple feeding points of the same inputs signal on the antenna; to our knowledge, this is the first dynamic antenna system achieving directional modulation with a single feeding port. 
%, the presented antenna operates onThis is a major difference from others works of directional modulation since they use arrays or multiport antennas for directional modulation. 
The antenna relies on simple external dynamic balun to achieve directional modulation and can therefore potentially be used in a ``black box'' fashion with existing antennas, and may therefore provide an easily added layer of security in future communications systems. 

\ifCLASSOPTIONcaptionsoff
  \newpage
\fi

\bibliographystyle{./bibliography/IEEEtran}
\bibliography{./bibliography/Dynamic_Antenna.bib,./bibliography/IEEEabrv.bib,./bibliography/IEEEexample.bib}

\end{document}